\documentstyle[bulk]{article}

\renewcommand{\textwidth}{16.5cm}

\newcommand{\kt}{k_t}

\newcommand{\xt}{x_t}
\newcommand{\nt}{n_t}
\newcommand{\ntdot}{\dot{\nt}}
\newcommand{\nzero}{n_0}

\newcommand{\Dsmall}{D_{\rm small}}

\newcommand{\rhot}{\rho_t}
\newcommand{\rhotthree}{\rho_t^{(3)}}
\newcommand{\rhothree}{\rho^{(3)}}

\newcommand{\rone}{{\bf r}_1}
\newcommand{\rtwo}{{\bf r}_2}
\newcommand{\rthree}{{\bf r}_3}
\newcommand{\r}{{\bf r}}
\newcommand{\qt}{q_t}
\newcommand{\Gsep}{G^{\rm sep}}
\newcommand{\St}{S_t}
\newcommand{\mumax}{\mu^{\rm max}}
\newcommand{\Qhat}{\widehat{Q}}
\newcommand{\Qstar}{Q^*}

\newcommand{\ta}{t_a}
\newcommand{\tstartwo}{t_{2}^{*}}
\newcommand{\tstarmany}{t_{m}^{*}}
\newcommand{\tstarmanyR}{t_{m,R}^{*}}
\newcommand{\tl}{t_l}
\newcommand{\te}{t_{\rm e}}
\newcommand{\tb}{t_{\rm b}}

\newcommand{\tc}{t_{\rm c}}

\newcommand{\Sdil}{S$^{\rm dil}$}
\newcommand{\Sconc}{S$^{\rm conc}$}

\newcommand{\Ne}{N_{\rm e}}
\newcommand{\re}{r_{\rm e}}
\newcommand{\rb}{r_{\rm b}}
\newcommand{\varphimax}{\varphi^{\rm max}}

\newcounter{fignumber}

\begin{document}



\renewcommand{\thepage}{}

\titleben{\LARGE \bf Reaction Kinetics in Polymer Melts}

\author{\Large 
BEN O'SHAUGHNESSY\ $^{1}$ \ and \ DIMITRIOS VAVYLONIS\ $^2$ \\ 
}

\maketitle

\ \\ \newline
{\large \center
$^1$ Department of Chemical Engineering \\
Columbia University\\ 
500 West 120th Street \\
New York, NY 10027, USA \\
e-mail: bo8@columbia.edu\\
} 

\vi

{\large \center
$^2$ Department of Physics \\ 
Columbia University \\
538 West 120th Street \\
New York, NY 10027, USA\\
e-mail: dvav@phys.columbia.edu\\
}

\  \newline


\pagebreak


\pagenumbering{arabic}

\large

\section*{ABSTRACT}

We study the reaction kinetics of end-functionalized polymer chains
dispersed in an unreactive polymer melt.  Starting from an infinite
hierarchy of coupled equations for many-chain correlation functions, 
a closed equation is derived for the 2nd order rate constant $k$ after
postulating simple physical bounds.  Our results generalize previous
2-chain treatments (valid in dilute reactants limit) by Doi
\citeben{doi:inter2}, de Gennes
\citeben{gennes:polreactionsiandii}, and Friedman and O'Shaughnessy
\citeben{ben:interdil_all_aip}, to arbitrary initial reactive group density
$n_0$ and local chemical reactivity $Q$.
Simple mean field (MF) kinetics apply at short times, $k \sim Q$.
For high $Q$, a transition occurs to diffusion-controlled
(DC) kinetics with $k \approx x_t^3/t$ (where $x_t$ is rms
monomer displacement in time $t$) leading to a density decay $n_t
\approx n_0 - n_0^2 x_t^3$.  If $n_0$ exceeds the chain overlap
threshold, this behavior is followed by a regime where $n_t \approx
1/x_t^3$ during which $k$ has the same power law dependence in time,
$k \approx x_t^3/t$, but possibly different numerical coefficient.  For
unentangled melts this gives $n_t \sim t^{-3/4}$ while for entangled
cases one or more of the successive regimes $n_t \sim t^{-3/4}$,
$t^{-3/8}$ and $t^{-3/4}$ may be realized depending on the magnitudes
of $Q$ and $n_0$.  Kinetics at times longer than the longest
polymer relaxation time $\tau$ are always MF.  If a DC regime has
developed before $\tau$ then the long time rate constant is $k
\approx R^3/\tau$ where $R$ is the coil radius.  We propose measuring
the above kinetics in a model experiment where radical end groups are
generated by photolysis.

\ignore{
For low $Q$ values this leads to a $\nt \twid 1/t$ regime.

($Q>\Qhat$ where $\Qhat \twid N^{-1/2}$ or $\Qhat \twid
\Ne N^{-3/2}$ for entangled and unentangled melts, respectively, where
$N$ is degree of polymerization and $\Ne$ the entanglement threshold)
} 

\vii

PACS numbers: 82.35.+t, 05.40.+j, 05.70.Ln \\

\ignore{
  \begin{benlistdefault}
    \item [82.35.+t]
 (Polymer reactions and Polymerization)
    \item [05.40.+j]
 (Fluctuation phenomena, random processes, and Brownian Motion)
    \item [05.70.Ln]
 (Nonequilibrium Thermodynamics, irreversible processes)   
  \end{benlistdefault}
} 

\ignore{
Possibly relevant PACS numbers: 

05.40.+j  Fluctuation Phenomena, random processes, and Brownian Motion  (*)
 
05.70.Ln  Nonequilibrium thermodynamics, irreversible processes (*)

05.20.-y  Statistical Mechanics

02.50.-r  Probability Theory, stochastic processes, and statistics

82.20.-w  Chemical Kinetics 

82.20.Db  Statistical theories (including transition state)

82.20.Mj  Nonequilibrium kinetics

82.35.+t  Polymer reactions and Polymerization (*)

82.40.-g  Chemical kinetics and reactions:special regimes and techniques 

}  

\pagebreak


\section{Introduction}

The study of polymer-polymer reaction kinetics is a fundamental
problem in polymer science.  Such reactions occur in many
technologically important processes such as vulcanization of rubbers
and free radical polymerization \citeben{flory:book}.  Compared to
analogous reactions between small molecules, polymer reaction kinetics
are novel in that they reflect static and dynamic properties of the
polymer chains which are host to the reacting groups.  Theoretical
studies predict that rate constants depend both on the degree of
polymerization and time
\citeben{doi:inter2,gennes:polreactionsiandii,ben:review_ijmp}.  Model
experiments 
\citeben{mitahorie:review,wisnudeltorkelson:inter_smallwins_best} and
numerical simulations
\citeben{grestkremerduering:crosslink_kinetics,grestkremerduering:crosslink_kinetics_and_relaxation}
testing these laws thus offer a way to probe fundamentals of polymer
dynamics .  In this paper we focus on reactions between
end-functionalized chains in a polymer melt.  So far, no systematic
experiment exists measuring rate constants in a melt as a function of
molecular weight and time.  It is our aim to motivate such model
experiments by developing a complete theoretical picture for such
reactions for direct comparison with experiment.

The situation we analyze is illustrated in fig. \ref{reactive_melt}.
We imagine that at $t=0$ a certain fraction of the chains randomly and
uniformly distributed in a monodisperse polymer melt carry chemically
reactive end-groups.  Experimentally this can be realized for example
by attaching photocleavable groups to the ends of a certain fraction
of the melt chains \citeben{ben:persistent}.  A laser pulse then
cleaves these groups into radical pairs (fig. \ref{melt_cleave}), one
of which is a small molecule and the other of which is attached to the
end of the chain (a macroradical) \citeben{turro:book}.  It was shown
in ref. \citeben{ben:persistent} that after a transient the small
radicals disappear, leaving behind a fraction of order unity of the
macroradicals.  The reaction kinetics of these {\em kinetically
isolated} macroradicals will then follow the kinetics developed in the
following.

Anticipating second order kinetics, our aim is to determine the
second order time-dependent rate constant $\kt$ defined by
                                                \begin{eq}{k}
\ntdot \equiv {d \nt \over dt} = -\kt \, \nt^2 \comma
                                                                \end{eq}
where $\nt$ is the number density of reactive groups at time $t$.  Now
for the usual case of reactions between small molecules, $\kt$ is
independent of time.  This is so because small molecules obey simple
Fickian diffusion, $\xt
\twid t^{1/2}$, where $\xt$ is the rms displacement in time $t$.  This
is a very dilute exploration of space (the exploration volume $\xt^3$
increases faster than $t$) and diffusion is fast enough to smooth out
any reaction-induced correlations
\citeben{gennes:polreactionsiandii,kangredner:segregation}.  Thus
reactant distribution remains uniform and random at all times, the
number of pairs per unit volume which are in contact at time $t$ is
therefore $a^3 \nt^2$, and hence $\ntdot \approx Q a^3 \nt^2$ where the
local reactivity $Q$ measures probability of reaction per unit time
when in contact.  This implies a time-independent rate constant $k
\approx Q a^3$.  For ``infinitely'' reactive groups for which $Q$ is
effectively equal to the rate at which diffusion brings apart a pair
of molecules in contact, \ie $Q\approx 1/\ta$ where $\ta$ is the
reactant relaxation time, this leads to the well known Smoluchowski
\citeben{smoluchowski:smallreactions} rate constant $k \approx a^3/\ta
\approx \Dsmall a$ where $\Dsmall \approx a^2/\ta$ is the small
molecule self-diffusivity.

Interpolymeric reaction kinetics are more complicated.  For times
shorter than the longest polymer relaxation time $\tau$, corresponding
to a diffusion distance of order the coil radius $R$, reactive
groups attached to polymer chains explore space in a dense way: The
dynamical exponent $z$, defined by $\xt \twid t^{1/z}$, is 4 or 8
depending on the degree of entanglement and time
\citeben{gennes:book,doiedwards:book}; thus the exploration volume
$\xt^3$ is increasing more slowly than $t$ and the number of times
reactive groups visit points within their exploration volumes is an
increasing function of time.  This was clearly pointed out by de
Gennes \citeben{gennes:polreactionsiandii} who differentiated between
``compact'' exploration when $z$ is greater than the dimensionality of
space $d$ and ``non-compact'' for $z<d$.

Now in the limit in which reactants are very dilute, $\nzero R^3 \ll
1$, the probability that by time $\tau$ a reactive group's exploration
volume contains a second reactive group is very small.  Hence for
$t<\tau$ reactions are due to the few isolated pairs which happened
to be initially within diffusive range,
\ie the members of which were initially within $\xt$ of one another. 
Due to the compact exploration, for times greater than $t$ the number
of collisions between such a pair grows without bound.  Therefore 
for ``infinitely'' reactive groups ($Q \ta \approx
1$) reaction is almost certain by $t$.  Thus a depletion hole of size $\xt$
starts to grow in the 2-body correlation function.  The number of
reacted pairs per unit volume is $\xt^3 \nzero^2$, which when
differentiated with respect to time implies a short time time-dependent
rate constant $\kt \approx \xt^3/t$.  Diffusion dynamics change at
$\tau$, since center of gravity Fickian diffusion of the whole chain
takes over, $\xt \twid t^{1/2}$.  The hole in the 2-body correlation
function stops growing and the situation is as in the small molecule
infinite reactivity case, but now with an effective capture
radius $R$ reflecting the short time compact dynamics.  Thus
                                                \begin{eq}{frog}
\kt \, \approx \, \casesbracketsii{\xt^3/t}{t<\tau}
                                  {R^3/\tau}{t>\tau}
\gap (Q \ta =1,\ n_0 R^3 \ll 1)
\period
                                                                \end{eq}
The $t>\tau$ expression for $k$ was first obtained by Doi
\citeben{doi:inter1,doi:inter2} for unentangled melts and was extended
by de Gennes \citeben{gennes:polreactionsiandii} to $t<\tau$ and
entangled cases.  Friedman and O'Shaughnessy
\citeben{ben:interdil_all_aip} derived $k$ from first principles using
renormalization group calculations.

The generalization of eq. \eqref{frog} to the concentrated regime,
$\nzero R^3 \gg 1$, is far less obvious and is one of the main
objectives of this paper.  In this case the timescale by which a
reactive group will almost certainly contain another group within its
exploration volume occurs during the compact regime, $t<\tau$.  During
this regime one can no longer assign reactions to pairs since a given
group has the opportunity to react with more than one partner.  A
proper study of the reaction kinetics requires a full many-body
treatment.  Such a many-body formalism involves an infinite hierarchy
of coupled dynamical equations for correlation functions of different
orders \citeben{wilemskifixman:dcreactionsgeneral,doi:reaction_secondquant2}.
Other than renormalization group analyses
\citeben{lee:aa_rg,leecardy:ab_rg}, this complication is typically
resolved by decoupling approximations expressing higher order
correlations in terms of lower order correlation functions
\citeben{kotominkuzovkov:book}.  In our
approach we are able to derive closed equations after assuming much
less restrictive bounds on the magnitude of correlation functions.  We
show that the implicit assumption of ref.
\citenum{gennes:polreactionsiandii}, namely that eq. \eqref{frog} is
also valid in the concentrated regime, is correct.

Our study is also general in terms of the local reactivity $Q$.  This
is very important experimentally, since typical chemical reactivities
are extremely small
\citeben{denisov:rateconstants_book,bernasconi:rateconstants_book}, $Q
\ta \lsim 10^{-6}$, with the exception of radical species which are
very close to being infinitely reactive, $Q \ta \approx 1$
\citeben{beckwith:radical_rateconstants_book}.

In the following section we present the many-body formalism describing
polymer-polymer reaction kinetics and introduce our assumptions on the
magnitude of correlation functions.  This results in a closed equation
for $k$.  We solve this equation in section 3 in terms of $\xt$ and
identify sequences of kinetic regimes depending on the values of $Q$
and $\nzero$.  We apply these general results to the case of
unentangled and entangled melts in sections 4 and 5, respectively.
Results are presented in the form of a ``phase-diagram'' in the
$Q$-$\nzero$ plane, different regions of which correspond to different
kinetic sequences.  We conclude with a discussion of our results in
section 6.


\section{Solution for Rate Constant $\kt$}

Consider the situation illustrated in fig. \ref{reactive_melt} in
which reactive ends are initially randomly distributed with density
$\nzero$.  Reactions commence at $t=0$.  Let us define
$\rhot(\rone,\rtwo)$ to be the 2-body correlation function of reactive
chain ends located at $\rone,\rtwo$.  Translational invariance implies
$\rhot$ is a function of $\rone$--$\rtwo$ only.  Now the reaction rate
per unit volume at point $\r$ is proportional to the number density of
pairs in contact at $\r$, $a^3 \rho(\r,\r)$, multiplied by the local
reactivity $Q$.  Here $a$ is the monomer size.  Using the fact that
translational invariance implies the average density $\nt$ is
independent of $\r$ one has
                                                \begin{eq}{onion}
\ntdot = - \lambda \, \rhot(0,0) 
\comma \gap \lambda \equiv Q a^3 
\period
                                                                \end{eq}
Thus the calculation of the reaction rate requires evaluation of the
2-body correlation function.  A rigorous dynamical equation for
$\rhot$ in the case of reactions between small molecules has been
derived by Doi in ref. \citenum{doi:reaction_secondquant2}.  In the
polymer reaction problem considered here, in order to derive a closed
relationship for $\rhot$ in terms of the degrees of freedom specifying
the location of the reactive ends only, one must first average out the
degrees of freedom specifying the locations of the other monomers;
this is non-trivial and requires renormalization group (RG) methods.
However, RG studies of 2-body bulk polymer reaction kinetics
\citeben{ben:intersemi_all,ben:interdil_all_aip} indicate that the basic physics is
completely captured by the approximate closing of the system in terms
of the coordinates of reactive groups: correct scaling behaviors are
obtained, only the prefactors being unreliable.  Making the
approximation of closure in terms of reactive degrees of freedom, then
the $\rhot$ dynamics involves the 3-body reactive groups correlation
function $\rhotthree(\rone,\rtwo,\rthree)$
\citeben{doi:reaction_secondquant2,ben:grosberg_book}:
                                                \begin{eqarray}{tomato}
\rhot(\rone,\rtwo) &=& \nzero^2 
- \lambda \int d\rone' \, d\rtwo' \int_0^t dt' \,
        G_{t-t'}(\rone,\rtwo,\rone',\rtwo')\, 
         \rho_{t'}(\rone',\rtwo') \delta(\rone'-\rtwo')
                                                        \drop
&-& \lambda \int d\rone' \, d\rtwo' \, d\rthree' \int_0^t dt' \,
        G_{t-t'}(\rone,\rtwo,\rone',\rtwo') \, 
        \rho_{t'}^{(3)}(\rone',\rtwo',\rthree')
        \curly{\delta(\rone'-\rthree') + \delta(\rtwo' - \rthree')}
\comma 
\drop
                                                          \end{eqarray}
where $G_t(\rone,\rtwo,\rone',\rtwo')$ is the {\em equilibrium} chain
end propagator giving the net weighting for two ends to arrive at
$\rone,\rtwo$ given starting points $\rone',\rtwo'$ in the absence of
reactions.  This is a well known object. 
\ignore{
(We neglect the change in $G$ due to the changing nonreactive polymer
matrix in the course of the reaction process.  This is reasonable
since products have degrees of polymerization of the same order as the
initial chains.)
} 
The sink terms on the right hand side of eq.
\eqref{tomato} (illustrated in fig.
\ref{sink_terms_bulk}) describe the three ways in which reactions
diminish $\rhot$ from its initial value $\nzero^2$.  The first {\em
two-body} sink term subtracts off pairs which failed to reach
$\rone,\rtwo$ because their members reacted with one another at
$\rone'$ at time $t'$.  The remaining two sink terms subtract off
pairs which would be at $\rone,\rtwo$ but only one member of which
reacted at time $t'$ at location $\rthree'$.  Such a reaction involves
a third chain, weighted by the appropriate 3-chain correlation
function.  These are {\em many-chain} terms; were they absent, one
would have a closed 2-chain system.  In refs. \citenum{doi:inter2} and
\citenum{gennes:polreactionsiandii} these terms were omitted. As
explained in the introduction one expects such a 2-body approach to be
valid in the dilute limit.  In eq. \eqref{tomato} we used a $\delta$
function as a reactive sink.  This is a coarse-grained description of
the reaction process over a scale of order the monomer size $a$.

Notice that eq. \eqref{tomato} is not closed in terms of $\rhot$ since
it involves the unknown $\rhotthree$.  It is in fact impossible to write a
closed exact equation for $\rhot$ since correlation functions of all
orders are coupled in an infinite hierarchy of dynamical equations
\citeben{doi:reaction_secondquant2,wilemskifixman:dcreactionsgeneral}.
This complication which arises in all many-body reacting systems is
typically resolved by approximating 3-body correlations in terms of
lower order correlation functions \citeben{kotominkuzovkov:book}.  In
this study we are able to write a closed equation for $\rhot$ after
assuming less restrictive bounds on the magnitude of correlation
functions.  To explore the constraints imposed by these bounds we
first transform eq.
\eqref{tomato} to an integral expression for the function $\qt(\r,0)
\equiv \rhot(\r,0)/\nt^2$.  Notice that eq.
\eqref{k} then implies $\kt = \lambda \qt(0,0)$.  It is  demonstrated in
appendix A that after making this change of variables, one has
                                                \begin{eqarray}{virtue}
\qt(\r,0) &=&  
          1 - \int d\rone' \int_0^t \,dt' G_{t-t'}(\r,0,\rone',\rone')\, k_{t'}
                                                \drop       
&+& 2 \int d\rone'\, d\rtwo' \int_0^t dt'\, G_{t-t'}(\r,0,\rone'+\rtwo',\rtwo')\,
k_{t'} \curly{\rho_{t'} (\rone'|0) - \rho_{t'}^{(3)} (\rone'|0,0)}
\comma
                                                \drop
                                                          \end{eqarray}
where $\rhot(\r|0)$ and $\rhotthree(\r|0,0)$ are the conditional reactive
chain end densities at $\r$, given one and two reactive ends,
respectively, at the origin:
                                                \begin{eq}{melon}
\rhot(\r|0) \equiv {\rhot(\r,0) \over \nt}
\comma \gap
\rhotthree(\r|0,0) \equiv {\rhotthree(\r,0,0) \over \rhot(0,0)}
\period
                                                                \end{eq}
Eq. \eqref{virtue} can be reexpressed in terms of the propagator 
$\Gsep_t(\rone,\rtwo) \equiv \int d\rtwo'\,
G_t(\rone',\rone'+\rone,\rtwo',\rtwo'+\rtwo)$, namely the probability
density two chains ends are separated by $\rone$ at time $t$ given
initial separation $\rtwo$.  Thus eq. \eqref{virtue} becomes
                                                \begin{eq}{myth}
\qt(\r,0) = 
 1 -  \int_0^t dt'\, \Gsep_{t-t'}(\r,0) \, k_{t'} \ + \ \varphi_t(\r) 
\comma
                                                                \end{eq}
where
                                                \begin{eqarray}{camel}
\varphi_t (\r) &\equiv&
\int_0^t dt' \int d\r' \, \Gsep_{t-t'}(\r,\r')\, k_{t'}\, \mu_{t'}(\r')
\comma \gap
\mu_{t}(\r) \, \equiv \, 2 \curly{
\rhot(\r|0) - \rhotthree(\r|0,0) 
} \period \drop
                                                                \end{eqarray}

We want to solve eq. \eqref{myth} for $\kt$ or equivalently for
$\qt(0,0)$.  This requires information on the properties of
$\varphi_t(0)$ which in turn involves 2-body and 3-body conditional
densities.  Now we make the assumption that the more reactive groups
placed at the origin, the lower the conditional density.  Chemical
reactivity can only induce anticorrelations.  Thus:
                                                \begin{eq}{assumption}
\rhotthree(\r|0,0)\, \leq\, \rhot(\r|0)\, \leq \nt
 \gap \mbox{(assumption)}
\period
                                                                \end{eq}
Eq. \eqref{assumption} immediately implies the following constraint
on $\mu$ (see eq. \eqref{camel})
                                                \begin{eq}{constraint1}
0 \, \leq \,\mu_t (\r)\, \leq \, 2 \nt \period
                                                                \end{eq}
Eq. \eqref{assumption} also implies that $\qt\le1$.  Thus the
magnitude of the second term on the rhs of eq. \eqref{myth} must
exceed that of the 3rd term, $\int_0^t dt' \Gsep_{t-t'} (\r,0) k_{t'}
\ge \varphi_t(\r)$.  This is the full inequality which must be obeyed
by the function $\mu_{t'}(\r')$ involving the unknown $\rhothree$.
Integrating over $\r$ yields the following less demanding
inequality:
                                                \begin{eq}{constraint2}
\int_0^t dt' k_{t'} \, \ge \, \int_0^t dt' \int d\r' \, k_{t'} \,
\mu_{t'} (\r') \period 
                                                                \end{eq}

On the strength of the above constraints, eqs. \eqref{constraint1} and
\eqref{constraint2}, we will now argue that the solution of
eq. \eqref{myth} for $\qt(0,0)$, but with the term $\varphi_t(0)$
deleted, gives the correct power law solution for $\kt$ to within
a constant prefactor.  Expressing $\qt(0,0)$ in terms of $\kt$,
setting $\r=0$ and deleting $\varphi_t(0)$, eq. \eqref{myth} becomes
                                                \begin{eq}{clock}
\kt = \lambda - \lambda \int_0^t dt' S_{t-t'} k_{t'} \comma
                                                                \end{eq}
where we have introduced the return probability $\St \equiv
\Gsep_t(0,0)$.  

In appendix B, a self-consistent argument is presented to justify the
deletion of $\varphi_t(0)$.  A summary of this argument is as follows.
Now if one accepts eq. \eqref{clock}, then eqs. \eqref{clock} and
\eqref{k} lead to a sequence of power law regimes in time, both for
$\kt$ and $\nt$ (these are explicitly obtained in the next sections).
Using these solutions in the constraints (eqs. \eqref{constraint1} and
\eqref{constraint2}) and in the expression for 
 $\varphi_t(0)$ (eq. \eqref{camel}), the function $\mumax_{t'}(\r')$
which maximizes $\varphi_t(0)$ (for a given time $t$) subject to the
constraints is determined.  This in turn implies an upper bound on
$\varphi_t(0)$, namely $\varphi_t^{\rm max}(0)$, according to eq.
\eqref{camel}.  It is then shown in appendix B that
 $\varphi_t^{\rm max}(0) \ll 1$.  (Of course, in reality there is one
unique value for $\varphi_t(0)$, since $\mu_{t'}(\r')$ is a determined
function, eq. \eqref{camel}.  What has been achieved here is to bound
this function given our incomplete knowledge.)  Thus $\varphi_t(0)$
may be deleted in eq. \eqref{myth} without error; even prefactors are
expected to be correct.  There is, however, one exception to this
simple state of affairs.  During those time regimes where $n_t\approx
1/\xt^3$ (see next sections) we find $\varphi_t^{\rm max}(0) = A$,
where $A$ is a constant of order unity.  For these regimes, we are
forced to make one further assumption, that $\kt$ remains a power law
during this regime.  In appendix B, it is shown that this implies that
the possibility of $\varphi_t(0)$ being of order unity may modify the
coefficient of $\kt$ only, not the power itself.

Thus in the following we simply solve eq. \eqref{clock}.  This
equation has been derived in ref. \citenum{gennes:polreactionsiandii}
starting from a 2-body formalism valid in the dilute reactive species
limit as discussed.  Here we argue that it is valid for all $\nzero$.
Expressions equivalent to eq. \eqref{clock} were the starting point in
ref. \citenum{oshanin:review} the authors of which pointed out the
existence of a $\nt\approx 1/\xt^3$ regime in polymer reaction
kinetics.

\ignore{
We use the following a self-consistent reasoning to argue for the
validity of eq. \eqref{clock}.  In the next section we solve eqs.
\eqref{clock} and \eqref{k} to obtain a series of successive power law
regimes in time, both for $\kt$ and $\nt$.  Then the
function $\mumax_{t'}(\r')$ which maximizes $\varphi_t(0)$ (for a
given time $t$) is determined, subject to the constraints in eqs.
\eqref{constraint1} and
\eqref{constraint2}.  This in turn gives an upper bound on
$\varphi_t(0)$, namely $\varphi_t^{\rm max}(0)$, when $\mumax$ is
substituted in eq. \eqref{camel}.  The calculation which is done in
Appendix B shows that during certain time regimes $\varphi_t^{\rm
max}(0) \ll 1$.  Thus $\varphi_t(0)$ may be deleted in eq.
\eqref{myth} without error (even the prefactors of the power law
solution are expected to be correct) during these time regimes.  The
only complication arises during a certain time regime during which the
integral term in eq. \eqref{myth} balances with the constant term 1.
During this regime $\varphi_t^{\rm max}(0) = A$, where $A$ is a
constant of order unity.  Assuming that $\kt$ is a still a power law
during this regime, we show in Appendix B that this implies that a
finite $\varphi_t(0)$ may only modify the coefficient of the power law
time dependence of $\kt$.  Thus in the following we simply solve eq.
\eqref{clock}.  This equation has been derived in ref.
\citeben{gennes:polreactionsiandii} starting from a 2-body formalism
valid in the dilute reactive species limit as discussed.  Here we
argue that it is valid for all $\nzero$.
} 

Now Laplace transforming $t \gt E$ one immediately solves eq.
\eqref{clock} for $k(E)$:
                                                \begin{eq}{citrus}
k(E) = {\lambda \over E (1+\lambda S(E))}
\comma
                                                                \end{eq}
which is the main result of this section.


\section{Kinetic Regimes and Timescales}

In the previous section we obtained a solution for $k$ in Laplace
space.  This involves the return probability $\St$.  It is easy to
show \citeben{gennes:polreactionsiandii} that $\St
\approx 1/\xt^3$ where $\xt$ is rms displacement.  In this section by 
substitution in eq. \eqref{citrus} we obtain solutions for $k$ and $n$
in terms of $\xt$.  Explicit forms of $\xt$ for unentangled and
entangled melts are considered in sections 4 and 5.

For times shorter than the longest polymer relaxation time $\tau$
(corresponding to $E \gg \tau$), $\xt \twid t^{1/z}$ with $z=4$ and
8 depending on time and the degree of entanglement
\citeben{gennes:book,doiedwards:book}.  Thus $\St \twid t^{-3/z}$
which implies $S(E) \twid E^{3/z-1}$ for $1/E \ll \tau$.  For times $t
\gg \tau$, center of gravity Fickian diffusion applies, $\xt \twid
t^{1/2}$.   Thus $\St \twid t^{-3/2}$ and it follows that for $1/E \gg
\tau$, $S(E)$ approaches its $E=0$ limit, $\int_0^\infty dt S(t)
\approx \tau/R^3$ (see refs. \citenum{doi:inter2,gennes:polreactionsiandii}).  
Thus, as a function of $1/E$, $S(E)$ increases and then saturates at
$\tau$.

To determine $\kt$ we must determine which of the two terms, $1$ or
$\lambda S(E)$ in eq. \eqref{citrus} is dominant.  There are two
cases.  If $Q$ (or $\lambda$) is sufficiently small, $Q<\Qhat$ (see
definition below), then $\lambda S(E) \ll 1$ for all $E$ values,
implying
                                                \begin{eq}{box}
\kt \approx \lambda \comma \gap  (Q<\Qhat) \period
                                                                \end{eq}
For $Q>\Qhat$ there exists a $1/E$ value (shorter than $\tau$),
corresponding to a timescale $\tstartwo$, after which $\lambda S(E) \gg
1$.  In this case during the regime $\tstartwo \ll E^{-1} \ll \tau$,
$k(E) \approx 1/E S(E) \twid E^{-3/z}$.  Thus Laplace inverting
eq. \eqref{citrus} we have:
                                                \begin{eq}{lemon}
\kt \approx \casesbracketsshortiii{\lambda}{t \ll \tstartwo}
                             {\xt^3/t}{\tstartwo \ll t \ll\tau}
                             {R^3/\tau}{t \gg\tau}
\comma \gap (Q>\Qhat) \period
                                                                \end{eq}
The quantities $\tstartwo$ and $\Qhat$ are defined by
                                                \begin{eq}{tstartwo}
{x^3_{\tstartwo} \over \tstartwo} = \lambda
\gap (\tstartwo<\tau) \comma \gap
\Qhat = {R^3 \over a^3 \tau} 
\period
                                                                \end{eq}
The value of $\tstartwo$ is determined by demanding continuity of
$\kt$.  In Laplace space, $\tstartwo$ corresponds to the $1/E$ value
value at which $\lambda S(E)$ becomes of order unity.  Eq.
\eqref{tstartwo} has a solution for $\tstartwo$ only if $Q$ is large
enough so that $\tstartwo<\tau$.  Thus the condition $\tstartwo=\tau$
defines $\Qhat$ in eq. \eqref{tstartwo}.

Now let us see what eqs. \eqref{box} and \eqref{lemon} imply for the
decay of $\nt$, considering the cases $Q>\Qhat$ and $Q<\Qhat$ in turn.

{\bf 1. Case $Q>\Qhat$.} The solution of eq. \eqref{k} is $\nt =
\nzero /( 1 + \nzero \int_0^t k_{t'} dt')$.  Thus using eq.
\eqref{lemon} one has
                                                \begin{eq}{mix}
{\nt\over \nzero} \approx \casesbracketsshortiii
                  {1/(1+t/\tstarmany)}   {t \ll \tstartwo}
                  {1/(1+ \nzero \, \xt^3)}  {\tstartwo \ll t \ll \tau}
                  {1/(1+t/\tstarmanyR)}  {t \gg \tau}
\gap (Q>\Qhat) \period
                                                                \end{eq}
We define the timescales $\tstarmany, \tstarmanyR$ and $\tl$ by
                                                \begin{eq}{olaj}
\tstarmany \equiv \inverse{\lambda \nzero} \comma \gap
\tstarmanyR \equiv {\tau \over \nzero R^3} \comma \gap
x^3_{\tl} = \inverse{\nzero}
\period
                                                                \end{eq}
Here $\tl$ is the time to diffuse a distance of the order of the
typical initial separation between reactants.  Notice that at a certain
timescale the time-dependent term in the denominator of $\nt$ in eq.
\eqref{olaj} becomes of order unity.  This time is one of $\tstarmany,
\tl$, and $\tstarmanyR$.  Different kinetic behavior occurs
depending on the ordering of these timescales amongst themselves and
with respect to $\tstartwo$ and $\tau$.

Let us consider first the kinetics
in the ``concentrated'' case, $\nzero R^3 >1$, \ie $\tl>\tau$.  Then
the monotonically increasing time-dependent term in the denominator of
$\nt$ in eq. \eqref{mix} becomes of order unity before $\tau$.  There
are two possibilities 
depending on whether or not $Q$ is large enough such that
$\tstartwo$ is smaller than $\tl$ (notice that $\tl$ is
independent of $Q$).  Now if indeed $\tstartwo<\tl$ one has
                                                \begin{eq}{sconc}
\nt \, \approx\,
\nzero - \lambda \nzero^2 t\
\stackrel{\tstartwo}{\ggt} \
\nzero - \nzero^2 \xt^3\
\stackrel{\tl}{\ggt} \
\inverse{\xt^3}\
\stackrel{\tau}{\ggt} \
\inverse{R^3\, t/\tau}\
\gap (\tstartwo<\tl<\tau,\ \mbox{\Sconc})
\period
                                                                \end{eq}
We use the symbol \Sconc\ for ``strong concentrated'' since eq. \eqref{sconc} applies
for high $Q$ and $\nzero$ values (see phase-diagrams of figs.
\ref{bulk_rouse}, \ref{bulk_rept}).  Notice that we do not need to specify
$Q>\Qhat$ in eq. \eqref{sconc} since only for such $Q$ values is
$\tstartwo<\tau$.   If on the other hand $\tstartwo>\tl$, one similarly has
                                                \begin{eq}{weak}
\nt \, \approx\,
\nzero - \lambda \nzero^2 t\
\stackrel{\tstarmany}{\ggt}\ 
\inverse{\lambda t}\
\stackrel{\tstartwo}{\ggt}\ 
\inverse{\xt^3}\
\stackrel{\tau}{\ggt} \
\inverse{R^3\, t/\tau}\
\gap (Q>\Qhat,\ \tl<\tstartwo,\ \mbox{W})
                                                                \end{eq}
where W stands for ``weak'' since these systems are less reactive when
compared to the strong concentrated regime (see figs.
\ref{bulk_rouse} and \ref{bulk_rept}).  Notice that
we do not need to specify $\tl>\tau$ in eq. \eqref{weak}, since this
is implied by $Q>\Qhat$ (or equivalently $\tstartwo<\tau$) and
$\tl<\tstartwo$.

Now in the the ``dilute'' case, $\nzero R^3 <1$, the time-dependent
term in the denominator of $\nt$ in eq. \eqref{mix} becomes of order
unity for $t>\tau$:
                                                \begin{eq}{sdil}
\nt \, \approx\,
\nzero - \lambda \nzero^2 t\
\stackrel{\tstartwo}{\ggt} \
\nzero - \nzero^2 \xt^3\
\stackrel{\tau}{\ggt} \
\nzero -  \nzero^2 R^3 {t \over \tau}\ \, 
\stackrel{\tstarmanyR}{\ggt} \
\inverse{R^3\, t/\tau}\
\gap
(Q>\Qhat,\ \tl>\tau,\ \mbox{\Sdil})
\period
                                                               \end{eq}
Here \Sdil\ stands for ``strong dilute.''

{\bf 2. Case $Q<\Qhat$.} Finally let us examine the $Q<\Qhat$ case.
Substituting eq. \eqref{box} in \eqref{k} one immediately has:
                                                \begin{eq}{ww}
\nt \, \approx\,
\nzero - \lambda \nzero^2 t\
\stackrel{\tstarmany}{\ggt} \
\inverse{\lambda t}\
\gap (Q<\Qhat,\ \mbox{WW)} 
\period
                                                                \end{eq}
We use the symbol WW (very weak) to refer to systems belonging to this
class which correspond to the lowest $Q$ values in which mean field
kinetics are always applicable.

We have thus identified 4 distinct behaviors, eqs.
\eqref{sconc}-\eqref{ww}, for the decay of $\nt$, corresponding to 4
regions in the $Q$-$\nzero$ plane.  In the following two sections we
explicitly plot these regions for both unentangled and entangled melts
after giving expressions for $\xt$ as dictated by the Rouse and
reptation model, respectively.


\section{Application to Unentangled Melts}

The dynamics of polymers in an unentangled melt, \ie with degree of
polymerization $N$ shorter than the entanglement threshold $\Ne$, are
well known to obey Rouse dynamics \citeben{doiedwards:book,gennes:book}: 
                                                \begin{eq}{rouse}
\xt \approx \casesbracketsii
{a \ (t/\ta)^{1/4}}{t < \tau}
{R \ (t/\tau)^{1/2}}{t > \tau}      
    \gap {\tau\over\ta}= N^2\comma\gap {R\over a}= N^{1/2}\gap
\period                                   
                                                                \end{eq}
Thus Rouse dynamics is characterized by compact ($z=4$) short time
behavior, followed by noncompact Fickian diffusion ($z=2$).

By substitution of eq. \eqref{rouse} in eqs. \eqref{tstartwo} and \eqref{olaj} 
we obtain explicit expressions for $\tstartwo$ and $\tl$:
                                                \begin{eq}{lizard}
\tstartwo = \ta/ (Q \ta)^{4}  \gap (Q>\Qhat) \comma \gap
\tl = \ta/(a^3 \nzero)^{4/3}  \gap (\nzero R^3 > 1)
                                                                \end{eq}
(the timescale $\tl$ is only relevant to the kinetics when $\nzero
R^3>1$).  Timescales $\tstarmany$, $\tstarmanyR$ are already given
explicitly in eq. \eqref{olaj}.  Using these expressions, in fig.
\ref{bulk_rouse} the lines $\tstartwo=\tau,
\, \tl=\tau$, and $\tl=\tstartwo$ have been drawn
in the $Q$-$\nzero$ plane.  Certain sections of these lines are
omitted, in those regions where they are irrelevant.  These lines
define four distinct regions.  It is straightforward to verify that
each one of these corresponds to one of the kinetic regimes defined in
eqs. \eqref{sconc}-\eqref{ww} as indicated in the figure.

Notice that the condition $\tl=\tstartwo$ also implies
$\tl=\tstartwo=\tstarmany$, as one can easily verify using eqs.
\eqref{tstartwo} and \eqref{olaj}.  In the $Q$-$\nzero$ plane, 
the $\tl=\tstartwo=\tstarmany$ line separating the \Sconc\ from the W
region thus defines a density-dependent reactivity $\Qstar$:
                                                \begin{eq}{qstar-rouse}
\Qstar \ta = (\nzero a^3)^{1/3} \comma 
                                                                \end{eq}
indicated in fig. \ref{bulk_rouse}. 

Perhaps the most interesting feature of the unentangled case is the
$\nt \twid t^{-3/4}$ decay of $\nt$ during the $\nt \approx 1/\xt^d$
regime in regions \Sconc, W.


\section{Application to Entangled Melts}

In the case of very long chains, $N>\Ne$, polymer motion is affected
by entanglements.  In the reptation model
\citeben{doiedwards:book,gennes:book} entanglements are assumed to
inhibit lateral chain motion on the scale of the ``tube'' of diameter
$\re = \Ne^{1/2} a$, corresponding to a portion of chain comprising
$\Ne$ units.  For times shorter than the diffusion time $\te = \Ne^2
\ta$ to distance $\re$, monomers do not feel the tube and obey
Rouse-like $t^{1/4}$ dynamics as in unentangled melts.  For $t>\te$,
the chain diffuses curvilinearly up and down the tube in 1-dimensional
$t^{1/4}$ Rouse motion.  During these ``breathing modes,'' monomer rms
displacement in space increases as $t^{1/8}$, since the tube is itself
a random walk.  The chain relaxes its configuration relative to the
tube by time $\tb= N^2\ta$, corresponding to monomer diffusion
distance $\rb = \re (N/\Ne)^{1/4}$.  For longer times, coherent
diffusion along the tube gives rise to $t^{1/4}$ rms monomer
displacement.  This regime persists until the
longest polymer relaxation or ``reptation'' time, $\tau = (R / \rb)^4
\tb = (N^3/\Ne)\ta$, by which time the chain has completely diffused
out of its initial tube into a new and uncorrelated one (here
$R=N^{1/2}a$).  The process then repeats itself indefinitely,
corresponding to long time Fickian center of gravity motion, $\xt = R
(t/\tau)^{1/2}$.  In summary,
                                                \begin{eq}{xtrept}
\xt\, \approx \casesbracketsshortiiii
{a \ (t/\ta)^{1/4}}      {t<\te\equiv \Ne^2\ta}
{\re \ (t/\te)^{1/8}}      {\te<t<\tb\equiv N^2\ta}
{\rb \ (t/\tb)^{1/4}} 
                     {\tb<t<\tau\equiv (R/\rb)^4= N^3\ta/\Ne}
{R\ (t/\tau)^{1/2}}                      {t>\tau}
                                                                \end{eq}
Thus, there are 3 compact regimes with a sequence of dynamical
exponents $z=4,8,4$, followed by noncompact $z=2$ exploration.

Substituting eq. \eqref{xtrept} in eq. \eqref{tstartwo} one obtains
                                                \begin{eq}{mailutha}
\tstartwo \approx \casesbracketsshortiii
        {\ta(Q\ta)^{-4}}            
                  {Q > \re^3/(\te a^3)}
        {\te(Q \te a^3/\re^3)^{-8/5}}
                   {\rb^3/(\tb a^3) < Q < \re^3/(\te a^3)}
        {\tb(Q\tb a^3/\rb^3)^{-4}}
                   {\Qhat< Q < \rb^3/(\tb a^3)}
                                                                \end{eq}
Notice that, unlike the unentangled case, the appropriate formula for
$\tstartwo$ is now dependent on the dynamical regime during which it
happens to occur.  This is determined by $Q$.  Similar remarks apply
to $\tl$, expressions for which depend on $\nzero$.  Eq.
\eqref{xtrept} in eq. \eqref{olaj} leads to
                                                \begin{eq}{bohr}
\tl \approx \casesbracketsshortiii
        {\ta(\nzero a^3)^{4/3}}
                       {\nzero>\re^{-3}}
        {\te(\nzero \re^3)^{8/3}}
                        {\re^{-3}>\nzero>\rb^{-3}}
        {\tb(\nzero \rb^3)^{4/3}}
                        {\rb^{-3}>\nzero>R^{-3}}
                                                                \end{eq}

Once again the condition $\tl=\tstartwo=\tstarmany$ defines a
density-dependent reactivity $\Qstar$:
                                                \begin{eq}{qstar-rept}
\Qstar = \casesbracketsshortiii
        {(\nzero a^3)^{1/3}/\ta}
                       {\nzero>\re^{-3}}
        {(\nzero \re^3)^{8/3}/(\te \nzero a^3)}
                        {\re^{-3}>\nzero>\rb^{-3}}
        {(\nzero \rb^3)^{4/3}/(\tb \nzero a^3)}
                        {\rb^{-3}>\nzero>R^{-3}}                
                                                                \end{eq}

Similarly to the unentangled case, the phase diagram of fig.
\ref{bulk_rept} is constructed.  Now the three $Q>\Qhat$ regions
develop fine structure.  In each of these subregions, crossovers in
$\nt$ behavior occur during different short time compact regimes. 

How does one use this phase diagram to determine the reaction kinetics
for a given system?  The system's $Q$ and $\nzero$ values define a
point in the diagram.  Depending on which of the 4 regions,
\Sconc,\Sdil, W, WW, this point happens to belong to, the reaction
kinetics are then given by the appropriate member of the 4 equations
\eqref{sconc}-\eqref{ww}.  But in these expressions for reaction
kinetics timescales $\tstartwo$ and $\tl$ may appear.  The
appropriate formulae for these timescales are determined by the
fine structure location of the point within the region (eqs.
\eqref{lizard} and \eqref{bohr}).  The other timescales, $\tstarmany,
\tstarmanyR$, if relevant, are given by eq. \eqref{olaj}. 

For example, let us consider point $\alpha$ in fig. \ref{bulk_rept}
belonging to region \Sconc.  According to eq. \eqref{sconc} the
relevant timescales (in addition to $\tau$) are $\tstartwo$ and $\tl$.  Fig.
\ref{bulk_rept} then indicates $\tstartwo <\te$ and $\te < \tl <\tb$.
Thus the $\nzero - \lambda \nzero^2 t$ regime of eq. \eqref{sconc}
occurs before $\te$, followed by $\nzero - \nzero^2 \xt^3$ which
occurs during the latter part of the first $\xt \twid t^{1/4}$ regime
and the earlier part of of the $\xt \twid t^{1/8}$ regime.  The
crossover to $\nt \approx 1/\xt^3$ occurs during the $t^{1/8}$ regime,
lasting up to $\tau$.  During these times $\nt \twid t^{-3/8}$
followed by $\nt \twid t^{-3/4}$.


\section{Conclusions}

This study has addressed the kinetics of reactions between
end-functionalized polymer chains diffusing in an unreactive polymer
melt matrix, as a function of initial reactant density $\nzero$ and chemical
reactivity $Q$.  Results were presented in the form of a ``phase
diagram'' in the $Q$-$\nzero$ plane.

At short times, simple mean field (MF) kinetics apply: the rate
constant $k \approx Q a^3$ is independent of time.  These kinetics
give rise to a density decay $\nt \approx \nzero - \nzero\,
t/\tstarmany$.  In regions W and WW of the $Q$-$\nzero$ plane (figs.
\ref{bulk_rouse} and \ref{bulk_rept}) these MF kinetics persist for
times greater than $\tstarmany$, leading to $\nt \twid 1/Q t$.
Physically, $\tstarmany$ as defined in eq. \eqref{olaj} is the
timescale after which a reactive group is very likely to have reacted
with the ``mean reactive field'' supplied by all other reactants in
the system.

A transition to diffusion-controlled (DC) kinetics occurs for
sufficiently reactive groups, $Q>\Qhat$.  This is a result of compact
exploration of space at times shorter than the longest chain
relaxation time $\tau$.  Thus a timescale $\tstartwo$ exists after
which any pair which was initially within diffusive range (\ie the
exploration volumes of its members are overlapping by $t$) is bound to
react.  If $\tstartwo$ occurs at times shorter than the time $\tl$
needed for a reactive end to diffuse a distance of order the typical
separation between reactive groups (regions \Sconc, \Sdil\ of figs.
\ref{bulk_rouse} and \ref{bulk_rept}), then the depletion in the
initial density $\nzero$ is proportional to the number of pairs with
initial separation less than $\xt$: $\nt \approx \nzero - \nzero^2
\xt^3$.  Equivalently, $\kt \approx \xt^3/t$
\citeben{gennes:polreactionsiandii}.  For unentangled melts this leads
to $\kt \twid t^{-1/4}$, while for entangled cases successive regimes
$\kt \twid t^{-1/4}$, $t^{-5/8}$ and $t^{-1/4}$ may exist depending on
$Q$, $\nzero$.

In regions \Sconc\ and W kinetics are diffusion-controlled for times
longer than $\tl$.  During these times the exploration volumes of
reactive groups overlap.  We find a density decay $\nt \twid 1/\xt^3$
\citeben{oshanin:review}.  Roughly, this means that only one reactant exists within
a region of size comparable to the exploration volume $\xt^3$; had
there been several, they would have reacted.  For unentangled melts
this implies $\nt \twid t^{-3/4}$, while an extra $\nt \twid t^{-3/8}$
regime arises in entangled systems.  These DC kinetics are analogous
to the anomalous long time decay $\nt
\approx 1/\xt^d$ in the case of small molecule reactions ($A+A \gt
\emptyset$) obeying Fickian dynamics, $\xt \twid t^{1/2}$, in spatial
dimensions below a critical dimension $d_c=2$.  For arbitrary
dynamics, $\xt \twid t^{1/z}$, the critical dimension is $d_c=z$; thus
$d=3$ is below $d_c$ for $t<\tau$ since $z$ is then either 4 or 8.
Notice that the rate constant during these times, $\kt
\approx \xt^3/t$, is the same as for short time ($t<\tl$) DC kinetics (as
described above).  However there is no obvious reason to expect them
to be identical since each one has different physical origins.
Indeed, our calculations suggest that the numerical coefficients of
these two rate constants are different.

All diffusion-controlled kinetics are truncated at $\tau$ which marks
a crossover to Fickian center of gravity polymer diffusion.  Thus long
time kinetics are always MF.  In cases where DC kinetics preceded, the
short time kinetics are reflected in a ``renormalized'' rate constant
$k \approx R^3/\tau$ \citeben{doi:inter2,gennes:polreactionsiandii}
leading to $\nt \approx 1/(R^3 t/\tau)$ at long times.

Experimentally, the most interesting region of the $Q$-$\nzero$ plane
is \Sconc.  This region includes the rather peculiar $\nt
\approx 1/\xt^3$ decay.  Probing reaction kinetics in this region
would require extremely reactive species such as radicals (or
electronically excited groups as in photophysical systems
\citeben{mitahorie:review}) since typical chemical reactivities, $Q
\ta \lsim 10^{-6}$, are far below the $Q=\Qhat$ threshold, even for
very long chains (see figs. \ref{bulk_rouse} and fig.
\ref{bulk_rept}).  One can imagine experimentally probing the \Sconc\ region by
photocleaving functional groups attached to polymer chain ends with a
laser pulse, generating radical pairs with initial density $\nzero$ as
in fig \ref{melt_cleave}.  Each pair consists of a polymer radical (P)
and a monomeric radical (M).  An experiment of this type was studied
theoretically in refs. \citenum{ben:persistent} and
\citenum{ben:persistent_letter}.  There are 2 underlying principles:
(1) For large $N$, since polymers are much less mobile than monomers,
the polymer-polymer reaction constant $k_{PP}$ is much less than the
monomer-monomer constant $k_{MM}$.  (2) Polymer-monomer reactions are
dominated by the more mobile monomer \citeben{ben:nm}, \ie $k_{MP}
\approx k_{MM}$ to within a prefactor of order unity.  In the
following we take $k_{MP}=k_{MM}$ for simplicity.  It was shown in
refs. \citenum{ben:persistent,ben:persistent_letter} that after a time
$T \approx (k_{MM} \nzero)^{-1}$ virtually all M have reacted, leaving
behind a finite fraction of the P: the polymer radicals are {\em
kinetically isolated} and the $t>T$ kinetics involve polymer-polymer
reactions only (fig. \ref{reactive_melt}).  Specifically, after time
$T$ a fraction $1/e$ of the P remain whilst only a fraction $\epsilon/e$ of
the M remain, where $\epsilon \equiv k_{PP}/k_{MM} \ll 1$.

Returning to the present case, if we assume all radical groups have
similar reactivities $Q \approx \ta^{-1}$ we identify $T \approx
(\lambda \nzero)^{-1} = \tstarmany$.  (This is the MF timescale we
have met previously, as expected: small molecules in $d=3$ obey MF
kinetics.)  Let us suppose $\nzero$ is chosen so the polymer radicals
are strongly overlapping, $\nzero R^3 \gg 1$ as in fig.
\ref{melt_cleave}.  The conclusion is that after $\tstarmany$ our
reactive polymer system belongs to the \Sconc\ region
\citeben{bulk:note} since $Q$ exceeds the $\Qstar$ threshold.  But in
region \Sconc\, by definition $\tstarmany$ is much less than the time
$\tl$ for a reactive polymer end group to diffuse the distance between
reactive ends.  Thus (consistent with the above arguments)
macroradicals cannot have substantially reacted with one another by
time $\tstarmany$.  Monitoring the decay of macroradical density for
$t>\tstarmany$, the full sequence of kinetics is thus experimentally
accessible, \ie $\nt
\approx \nzero - \nzero^2 \xt^3$ for $\tstarmany < t < \tl$ followed
by $\nt \approx 1/\xt^3$ for $\tl<t<\tau$.  Experiments of this type
thus hold promise for the measurement of fundamental scaling laws of
polymer-polymer reaction kinetics.

\nopagebreak

\ignore{  ORIGINAL VERSION IN 1st SUBMISSION
One can imagine probing the \Sconc\ region
by photocleaving chemical groups attached to polymer chain ends with a
laser pulse.  Indeed, suppose the small radical - macroradical pairs
are generated with sufficiently high density so that reactive chains
are overlapping.  Radicals are extremely reactive ($Q
\ta \approx 1$) and are therefore above the $\Qhat$ threshold.  Now
reactions between small molecules and between small molecules and
macroradicals obey simple MF kinetics (see ref. \citenum{ben:nm}).
Hence most small radicals will have reacted by a time of order
$\tstarmany$.  But in region \Sconc, $\tstarmany$ is much {\em
shorter} than the time $\tl$ needed for a reactive end-group to
diffuse a distance of order the typical separation between reactive
ends.  Thus macroradicals cannot have reacted with one another.  In
ref. \citeben{ben:persistent} it was shown that in fact during
this early regime when the small radicals disappear, a fraction of
order unity of them react with one other, and a fraction of order
unity with macroradicals.  The conclusion is that a fraction of order
unity of the macroradical population will remain unreacted long after
the small radicals disappear and long before the transition to the
$\nt \approx 1/\xt^3$ regime occurs.  Experiments of this type thus
hold promise for the measurement of fundamentals of polymer-polymer
reaction kinetics.
} 


\vspace{.1in}

This work was supported by the National Science Foundation under grant
no. DMR-9403566.  


{\appendix

\section{Derivation of eq. (5)}

Eq. \eqref{tomato} can be written equivalently as
\citeben{doi:reaction_secondquant2}
                                                \begin{eq}{diffusion}
{\partial \rhot(\rone,\rtwo) \over \partial t}\,  -\,  {\cal D}\,
\rhot(\rone,\rtwo)
= - \lambda \rhot(\rone,\rtwo) \delta(\rone-\rtwo)
  - \lambda \int d\rthree\, \rhotthree(\rone,\rtwo,\rthree) 
   \curly{\delta(\rone- \rthree) + \delta(\rtwo-\rthree)}
                                                                \end{eq}
where the terms on the right hand side represent reaction sinks and
${\cal D}$ is the diffusion operator.  The propagator $G$ in eq.
\eqref{tomato} is the
inverse of $(\partial/\partial t - {\cal D})$.  Notice that $-{\cal
D}\rho$ is equal to the divergence of the reactive chain end current
${\bf j}$ at $\rone,\rtwo$: ${\cal D}\rho = - (\nabla_{\rone} +
\nabla_{\rtwo}) {\bf j(\rone,\rtwo)}$.  (For example for small
molecules one would have ${\bf j} = - D (\nabla_{\rone} +
\nabla_{\rtwo})\rho$, where $D$ is the diffusion coefficient.)  Since
${\bf j}$ must be linear in $\rho$, one has ${\cal D}\rhot = {\cal
D}\qt
\nt^2 = \nt^2 {\cal D} \qt$.  Then, substituting $\rhot(\rone,\rtwo)
= \qt(\rone,\rtwo) \nt^2$ in eq. \eqref{diffusion} one obtains
                                                \begin{eqarray}{shell}
{\partial \qt(\rone,\rtwo) \over \partial t}\,  -\,  {\cal D}\,
\qt(\rone,\rtwo)
= - \lambda \qt(\rone,\rtwo) \delta(\rone-\rtwo)
   + 2 \rhot(\rone-\rtwo,0) \kt \nt
  - 2 \rhotthree(\rone-\rtwo,0,0)/\nt^2 \drop
                                                                \end{eqarray}
where we have performed the $\rthree$ integration in eq.
\eqref{diffusion} and used the identities $\rhot(\rone,\rtwo) = \rhot(\rone-\rtwo,0)$ and
$\rhotthree(\rone,\rtwo,\rone)=\rhotthree(\rone,\rtwo,\rtwo)=\rhotthree(\rone-\rtwo,0,0)$
which result from translational invariance.  Inverting eq.
\eqref{shell} one obtains eq. \eqref{virtue} after using eqs.
\eqref{k}, \eqref{onion} 
and the definitions of the conditional densities in eq. \eqref{melon}.


\section{Bounds on $\varphi_t(0)$}

This appendix shows that the deletion of the positive term
$\varphi_t(0)$, which results in eq. \eqref{myth} resulting in eq.
\eqref{clock} gives the correct scaling solution for $\kt$.  We show
this by first determining an upper bound, namely $\varphimax_t(0)$, by
finding the function $\mumax_{t'}(\r')$ in the definition of $\varphi$
( eq. \eqref{camel}), which subject to the constraints satisfied by
the real $\mu$, eqs. \eqref{constraint1} and
\eqref{constraint2}, maximizes $\varphi_t(0)$.  We consider each of
the 4 regions in figs. \ref{bulk_rouse}, \ref{bulk_rept} in turn. 

\subsubsection*{1. Region \Sconc}

{\em A. Times $t \ll \tl$.}   During these times $\nt \approx \nzero$
to within higher order corrections (see eq. \eqref{sconc}).  In this
case we just need to consider the constraint imposed on $\mu$ by eq.
\eqref{constraint1} only (including constraint \eqref{constraint2} would lead
to an even more strict bound which is not necessary here).  Thus eq.
\eqref{constraint1} implies $\mumax_{t'}(\r') = 2 \nt$.  Substituting
$\mumax$ in eq. \eqref{camel} and performing the $\r'$ integration
leads to
                                                \begin{eq}{chair}
\varphimax_t(0) = \nzero \int_0^t dt' \kt' \approx 
        \casesbracketsshortii {t/\tstarmany}  {t \ll \tstartwo}
                              {\nzero \xt^3}  {\tstartwo \ll t \ll \tl}
                                                                \end{eq}
after use of eq. \eqref{lemon}.  In the \Sconc\ regime
$\tstartwo\ll\tstarmany$ and $t \ll \tl$.  Thus eq. \eqref{chair} implies
$\varphimax_t(0) \ll 1$.  Hence during this regime $\varphi$ is
unimportant since it is much smaller than terms 1 and $\qt(0,0)$ in eq.
\eqref{myth}.  Eq. \eqref{clock} is thus expected to give the correct
$k$ (even the prefactor will not be modified by deletion of $\varphi$).

{\em B. Times $\tl \ll t \ll \tau$.} In this case we consider both
constraints \eqref{constraint1} and \eqref{constraint2}.  Now clearly
from its definition in eq. \eqref{camel}, $\varphi_t(0)$ is maximized
when the integral of $k(t') \mu_{t'}(\r')$ has its maximum value; this
corresponds to replacing the inequality in
\eqref{constraint2} by an equality.  Moreover, $\mu_{t'}(\r')$ should
be distributed around the points at which $\Gsep_{t-t'}(0,\r')$ in the
definition of $\varphi_t(0)$ in eq.
\eqref{camel} is maximum.  But $\Gsep$ has the well known scaling form 
                                                \begin{eq}{blue}
\Gsep_t (0,\r) \approx \casesbracketsshortii
                           {1/\xt^3 \ \ }{ r < \xt}
                           {0}           {r > \xt}
                                                                \end{eq}
Thus $\Gsep_{t-t'}(0,\r')$ is
roughly constant (in space) for $r' < x_{t-t'}$ and vanishes for
larger $r'$.  For a given $r'$, it achieves maximum values at the
maximum available $t'$ values outside the vanishing region.
Furthermore, we note that the maximum amplitude of $\mumax_{t'}$ is of
order $n(t')$.  Hence the function maximizing $\varphi_t(0)$ is
approximately
                                                \begin{eq}{worst-case}
\mumax_{t'}(\r') \approx 
\casesbracketsshortii  
                  {n(t')\, \Theta ( x_{t-t'} - r' )} {\tc<t'<t}
                  {0}                             {t'<\tc}
\comma \gap
\int_0^t k(t') dt' = \int_{\tc}^{t} dt' \int d\r' k(t') \mumax_{t'}(\r')
\comma
                                                                \end{eq}
where $\Theta$ is the step function ($\Theta(x)=0$ for
$x<0$,$\Theta(x)=1$ for $x\ge 0$).  That is, $\mumax_{t'}(\r')$ is
localized at $(t', \r')$ close to $(t, 0)$ where it has its maximum
amplitude.  The time $\tc$ defines the lower limit of the support of
$\mumax$ and is determined by demanding that the total integral of $k
\mumax$ yields the required value.  
Assuming that $\tc$ is of order $t$ (which is verified
self-consistently), we determine $\tc$ by substituting $\kt$ and $\nt$
from eqs. \eqref{lemon} and \eqref{sconc} in eq. \eqref{worst-case}
and performing the $\r'$ integration.  It is simple to show that this
implies $\int_{\tc/t}^1 (1-u)^{3/z}/u \, du \approx 1$, after using $\xt
\twid t^{1/z}$.  Hence $\tc = \beta\, t$, where $\beta$ is a constant
of order unity ($0<\beta<1$).  

Substituting $\mumax$ into the expression for $\varphi_t(0)$ in eq.
\eqref{camel} and integrating over $\r'$ one obtains
$\varphimax_{t}(0) \approx \int_{\tc}^{t} dt' k(t') n(t') \approx
\ln[n(\tc)/n(t)]$ after use of eq. \eqref{k}.  Since $\tc=\beta\,t$
thus $\varphimax_t(0) = A$, where $A$ is a constant of order unity.

We now argue that, assuming $\kt$ is still a power law, and given the
boundndess of $\varphi_t(0)$, this implies $\kt
\approx \xt^3/t$ during this regime.  Substituting $\kt \twid
t^{-\alpha}$ in the integral term in eq. \eqref{myth} and setting
$\r=0$, one easily derives the time-dependence of this term is
$t^{\alpha-(3-z)/z}$.  Since $\varphi_t(0)$ positive and bounded by
$A$, and since the term $\qt(0,0)$ tends to 0 at long times, the only
way to satisfy eq. \eqref{myth} for $\r=0$ is by setting
$\alpha=(3-z)/z$.  This implies $\kt
\approx \xt^3/t$.  For this value of $\alpha$ the integral term in eq.
\eqref{myth} tends to a constant.  Hence if $\varphi_t(0)$ is of order
unity during this regime, it must be a constant in order for eq.
\eqref{myth} to be satisfied.  The actual numerical prefactor of $\kt$
will then be a function of the value of $\varphi_t(0)$.

{\em C. Times $t \gg \tau$.} During these times $\nt \approx
1/(R^3 t/\tau)$ and $\kt \approx R^3/\tau$ (see eqs. \eqref{sconc} and
\eqref{lemon}).  We use the same reasoning as in part 1.B. up to eq.
\eqref{worst-case}.   As in 1.B., it is easy to show that $\tc/t = 1 -
\gamma$, where $\gamma \approx [R^3 t/(\xt^3 \tau)]^{2/5}$ and that
$\varphimax_t(0) \approx \gamma$.  Since $\gamma \ll 1$ for $t \gg
\tau$, it follows that $\varphi_t(0) \ll 1$ and that $\varphi_t(0)$ 
may be deleted from eq. \eqref{myth}.

\subsubsection*{2. Region W}

{\em A. Times $t \ll \tstarmany.$} Similar arguments as in part 1.A.
apply.

{\em B. Times $\tstarmany \ll t \ll \tstartwo.$} The procedure is the
exactly the same as in part 1.C., but now using $\kt \approx \lambda$
and $\nt \approx 1/\lambda t$ (see eqs. \eqref{lemon} and
\eqref{weak}).  In this case one has $\tc/t = 1 - \gamma$ with $\gamma
\approx (\lambda t/\xt^3)^{2/5}$ and $\varphimax_t(0) \approx \gamma$.
Since $\gamma \ll 1$ for $t \ll \tstartwo$ it follows that $\varphi_t(0)
\ll 1$ during this regime. 

{\em C. Times $t \gg \tstartwo.$} Identical to 1.B. for $\tstartwo \ll
t \ll \tau$ and to part 1.C. for $t \gg \tau$.  

\subsubsection*{3. Region \Sdil}

{\em A. Times $t \ll \tstarmanyR$.} For times $t \ll \tau$, same as
part 1.A. During this regime $\varphimax_t(0)$ in eq. \eqref{chair} is
much smaller than unity since $\tl>\tau$.  For $\tau \ll t \ll
\tstarmanyR$, similarly to 1.A., one finds $\varphimax_t(0) =
t/\tstarmanyR$ which is much smaller than unity.

{\em B. Times $t \gg \tstarmanyR$.} Same as part 1.C. 

\subsubsection*{4. Region WW}

{\em A. Times $t \ll \tstarmany$.} Same as part 1.A. 

{\em B. Times $t \gg \tstarmany$.} Same as part 2.B., but now notice
that $\gamma \ll 1$ for all $t$ ($\tstartwo$ is not defined in this
region).


}


\pagebreak


\begin{thebibliography}{10}
\bibitem{bulk:note}
A slight complication here is that during the transient $0<t<t^*_m$ the
  polymer-polymer reaction rate $k_{PP}(t)$ is time-dependent. However, the
  conclusions are essentially unchanged: the important feature is that
  $k_{PP}(t) \ll k_{MM}$ for all times $t \gg t_a$.
\end{thebibliography}


                     \begin{thefigures}{99}

\figitem{reactive_melt}

End-functionalized polymer chains (coil radius $R$) dispersed within a
polymer melt of inert but otherwise identical chains.  The
initial density of reactive groups is $\nzero$.  

\figitem{melt_cleave}

Schematic of the situation immediately after photolysis of functional
end-groups by a laser pulse generating radical pairs (consisting of a
monomeric radical and a macroradical).  The concentrated regime
(overlapping reactive coils) is illustrated.  After a transient most
of the more mobile monomeric radicals react leaving behind a fraction
of order unity of the polymer radicals.  The situation of fig.
\ref{reactive_melt} is then recovered.

\figitem{sink_terms_bulk}

The depletion in the number density of reactive groups at
$\rone,\rtwo$ originates from two terms in eq. \eqref{tomato}.  The
two-body term subtracts off those pairs which would have been at
$\rone,\rtwo$ at time $t$, but failed to do so because {\em both}
members reacted at an earlier time $t'$ at point $\rone'$.  The
many-body term subtracts off pairs which would have been at
$\rone,\rtwo$ had there been no reactions, but failed to arrive
because {\em one} member of the pair reacted at an earlier time.

\figitem{bulk_rouse}

(a) Unentangled melts, reaction kinetics ``phase diagram'' in $Q$-$\nzero$
plane.  Axes are logarithmic and units chosen such that $\ta=a=1$.
Maximum possible density is $\nzero a^3=1/N$ (all chains
functionalized).  Different sequences of kinetic behavior arise
in the 4 regions \Sconc, \Sdil,  W and WW.
(b) As (a), but reactivities and densities expressed in terms of
degree of polymerization N.

\figitem{bulk_rept}

As fig. \ref{bulk_rouse}, but for entangled melts.  Regions
\Sconc, \Sdil\ and W now develop fine structure.  In a given
sub-region each relevant timescale occurs within a given reptation
diffusion regime thus defining a unique sequence of kinetic regimes.

                     \end{thefigures}


\pagebreak

%
%
%
%
%

\begin{figure}[t]

\epsfxsize=4.5in  \epsffile{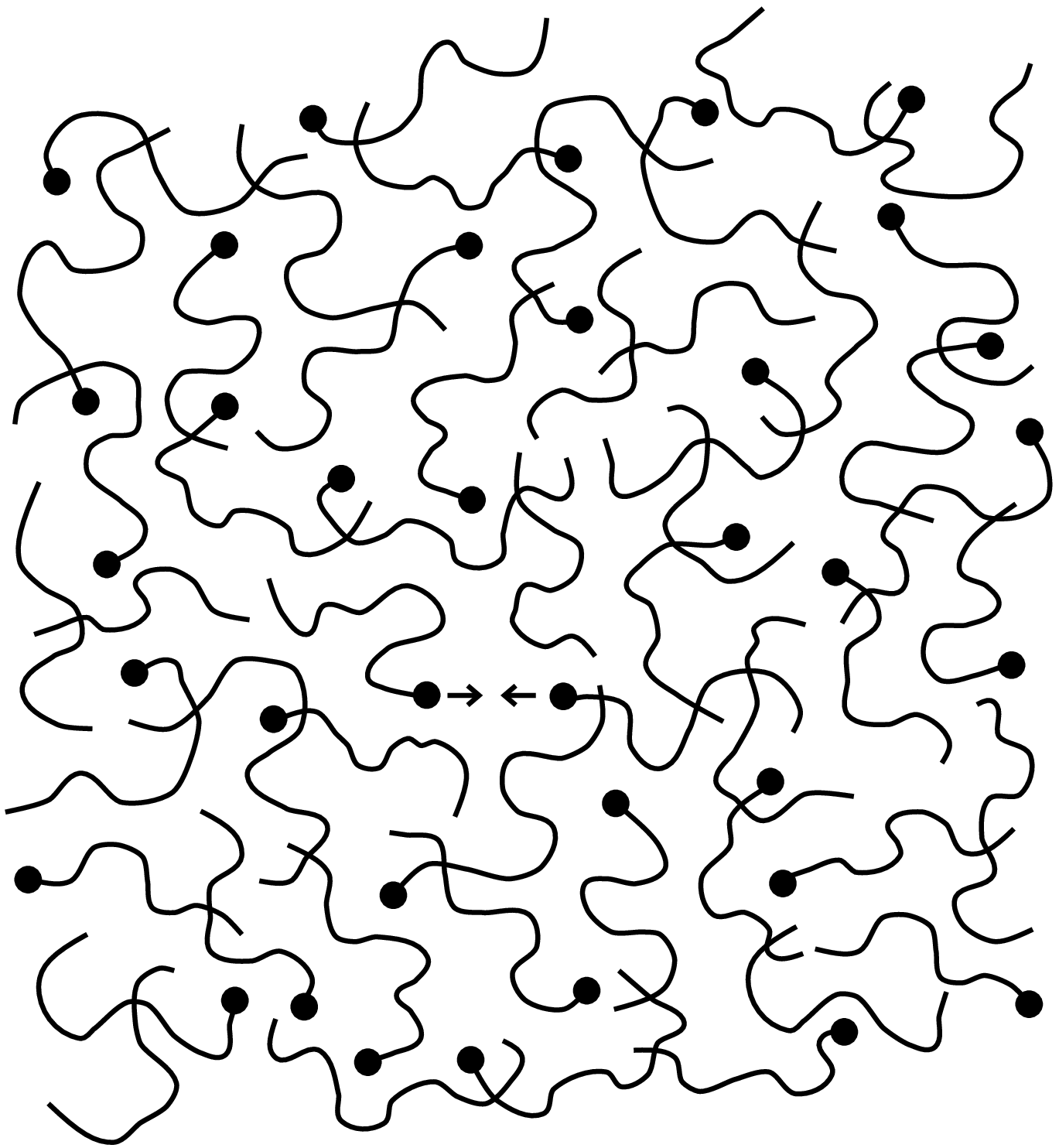}

\end{figure}

\mbox{\ }

\vfill

\addtocounter{fignumber}{1}
\mbox{\ } \hfill {\huge Fig.\@ \thefignumber} 

\pagebreak
\begin{figure}[t]

\epsfxsize=4.5in  \epsffile{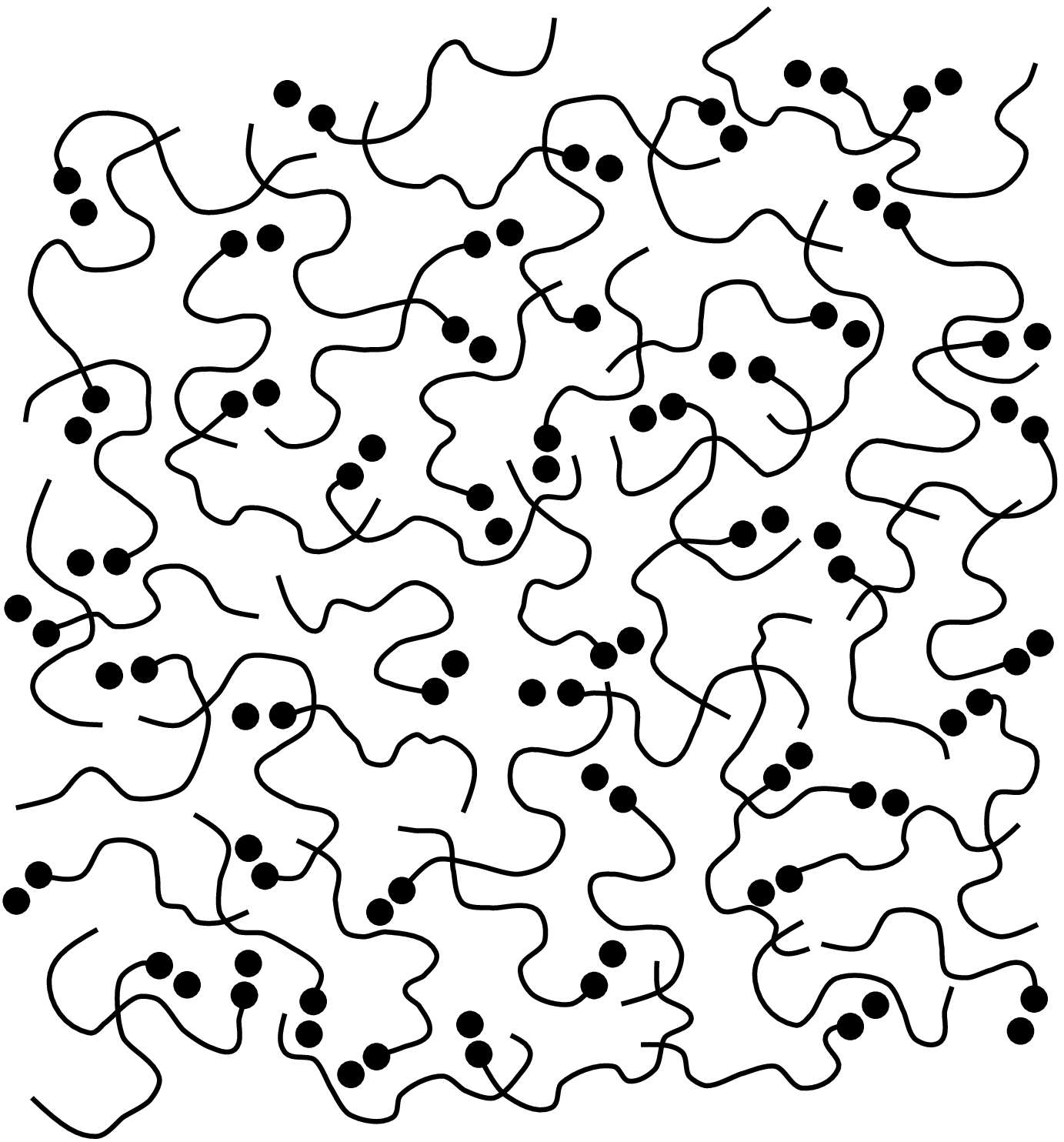}

\end{figure}

\mbox{\ }

\vfill

\addtocounter{fignumber}{1}
\mbox{\ } \hfill {\huge Fig.\@ \thefignumber} 

\pagebreak
\begin{figure}[t]

\epsfxsize=\textwidth \epsffile{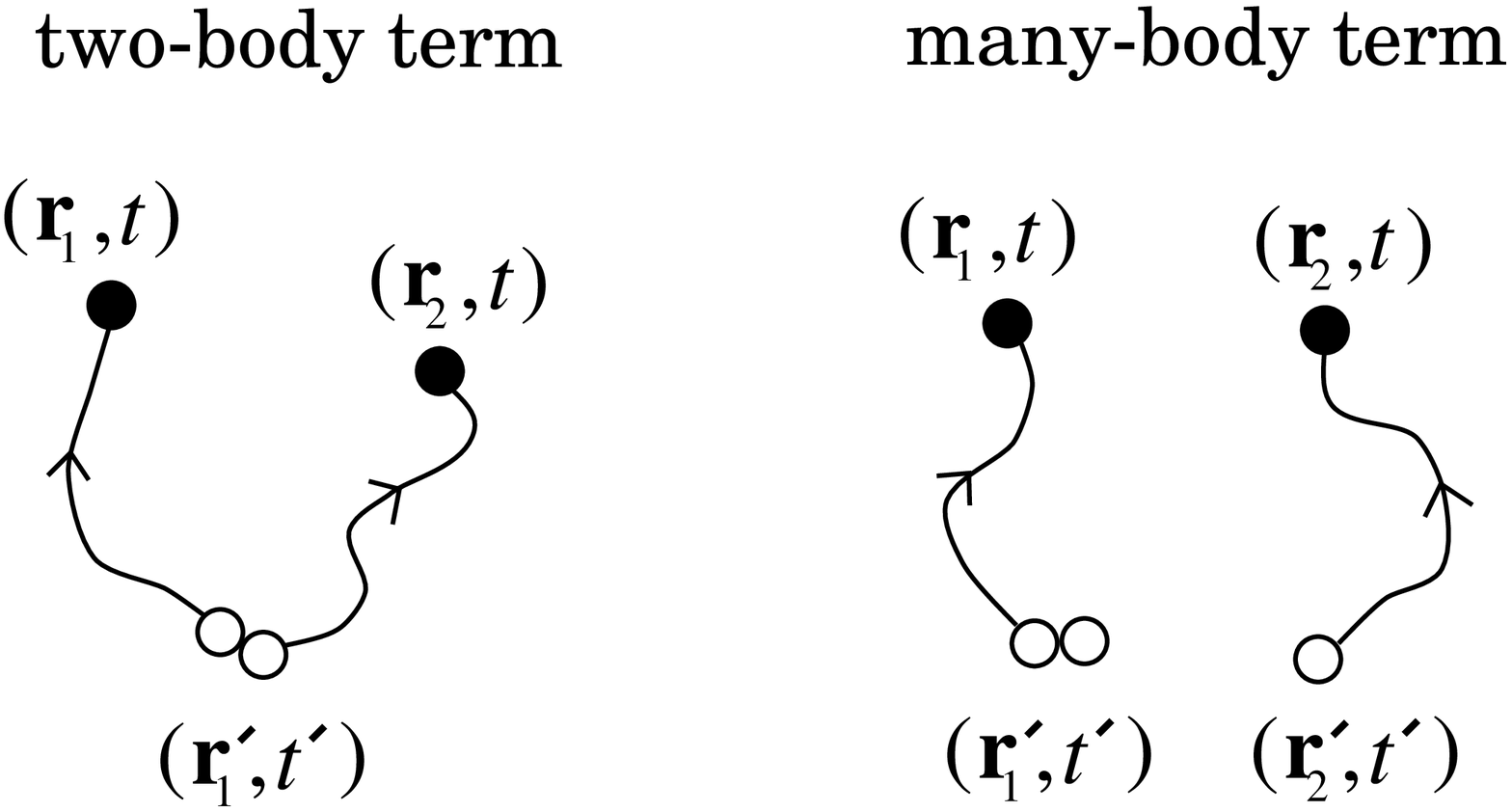}

\end{figure}

\mbox{\ }

\vfill

\addtocounter{fignumber}{1}
\mbox{\ } \hfill {\huge Fig.\@ \thefignumber} 

\pagebreak
\begin{figure}[t]

\epsfysize=5in \epsffile{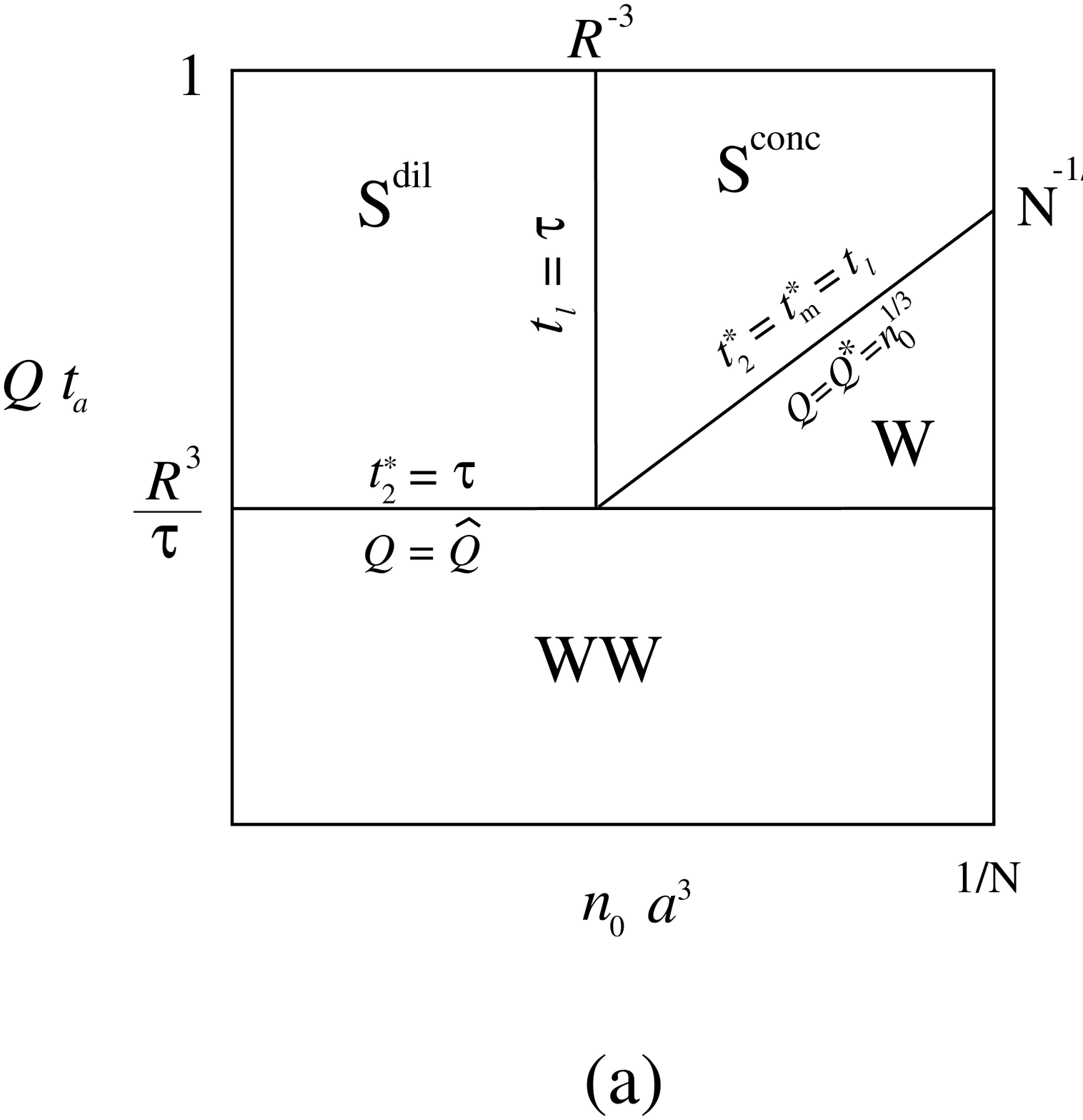}

\end{figure}

\mbox{\ }

\vfill

\addtocounter{fignumber}{1}
\mbox{\ } \hfill {\huge Fig.\@ \thefignumber (a)}

\pagebreak
\begin{figure}[t]

\epsfysize=5in \epsffile{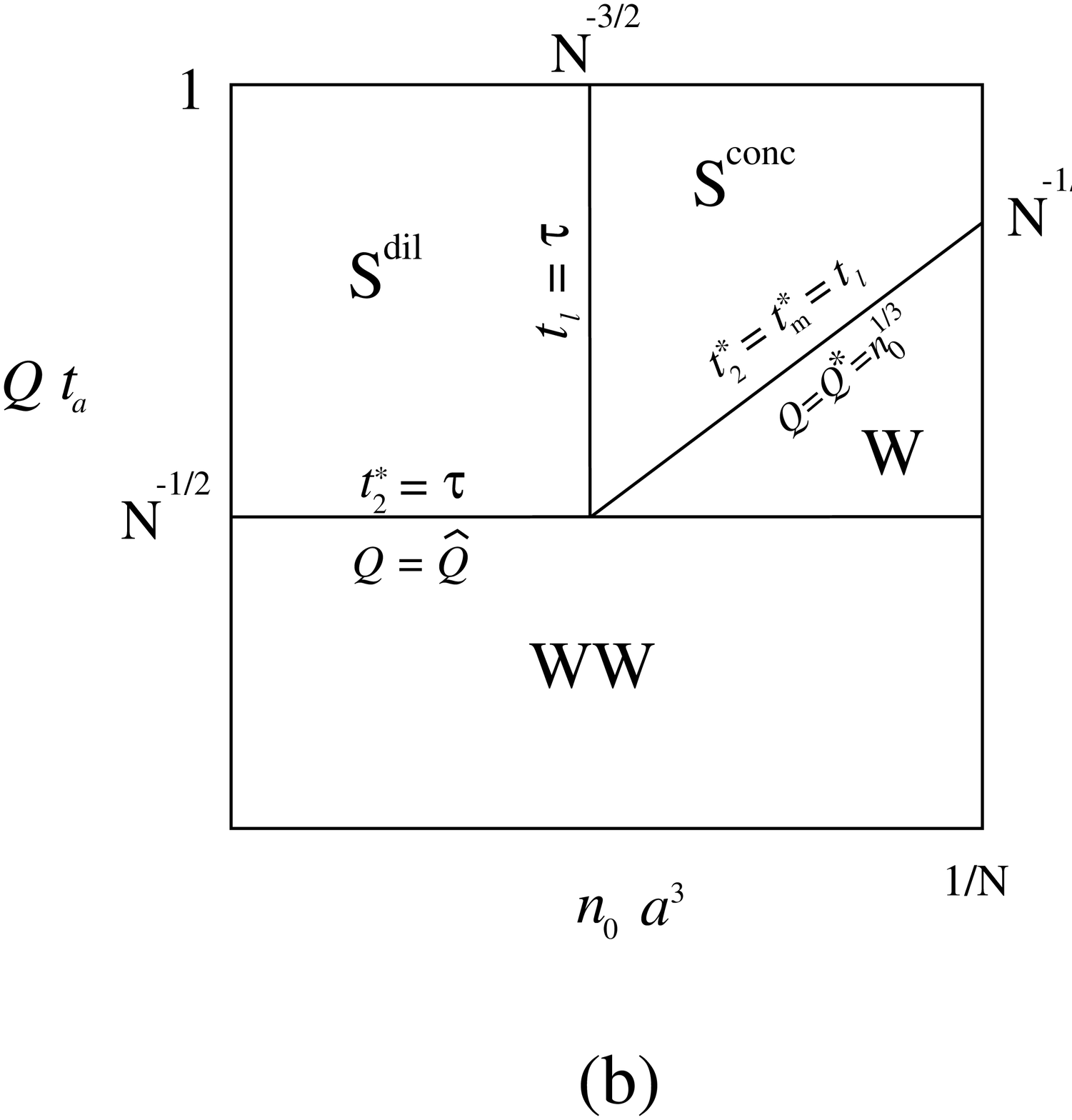}

\end{figure}

\mbox{\ }

\vfill

\mbox{\ } \hfill {\huge Fig.\@ \thefignumber (b)}

\pagebreak
\begin{figure}[t]

\epsfysize=5in \epsffile{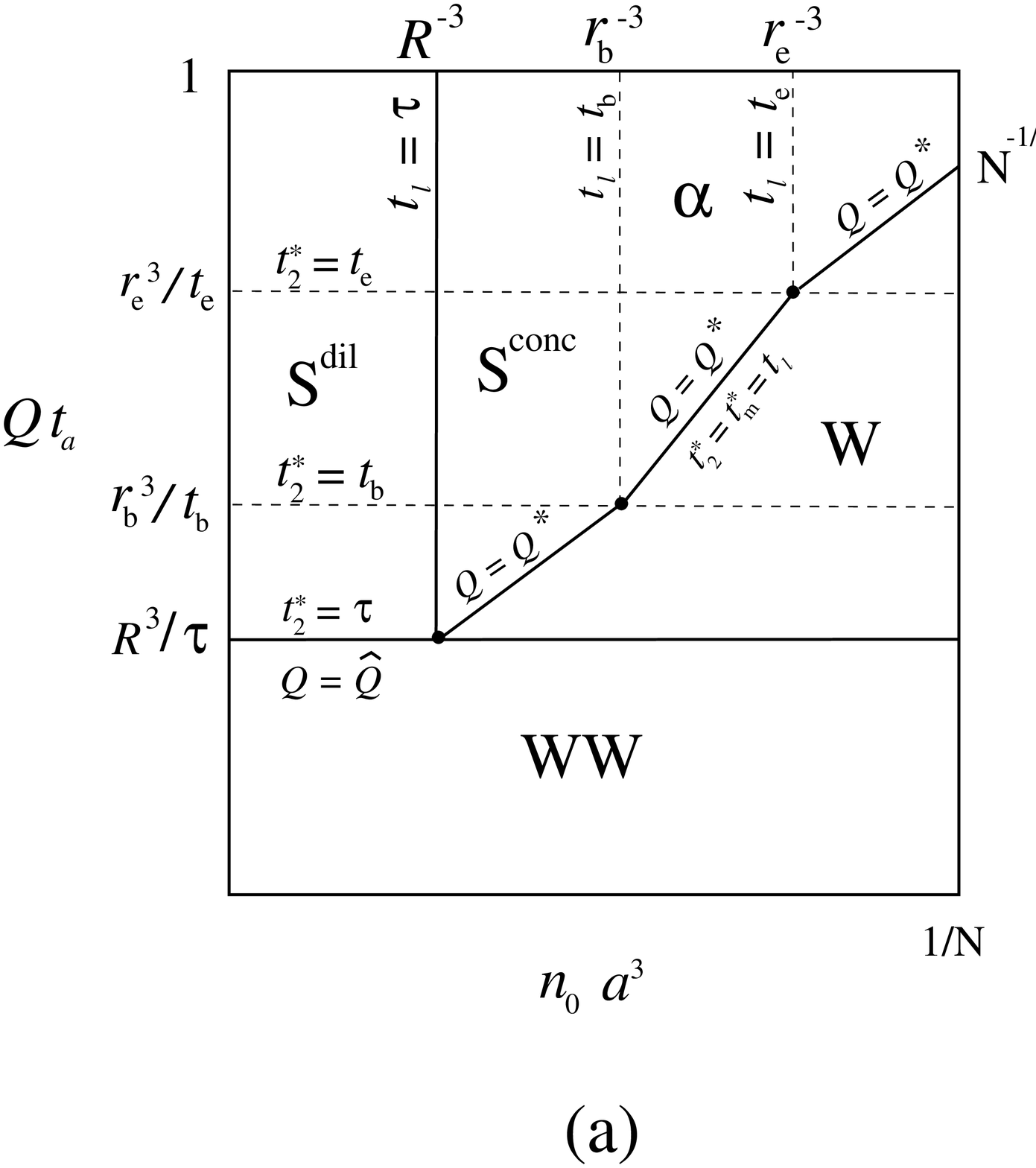}

\end{figure}

\mbox{\ }

\vfill

\addtocounter{fignumber}{1}
\mbox{\ } \hfill {\huge Fig.\@ \thefignumber (a)} 

\pagebreak
\begin{figure}[t]

\epsfysize=5in \epsffile{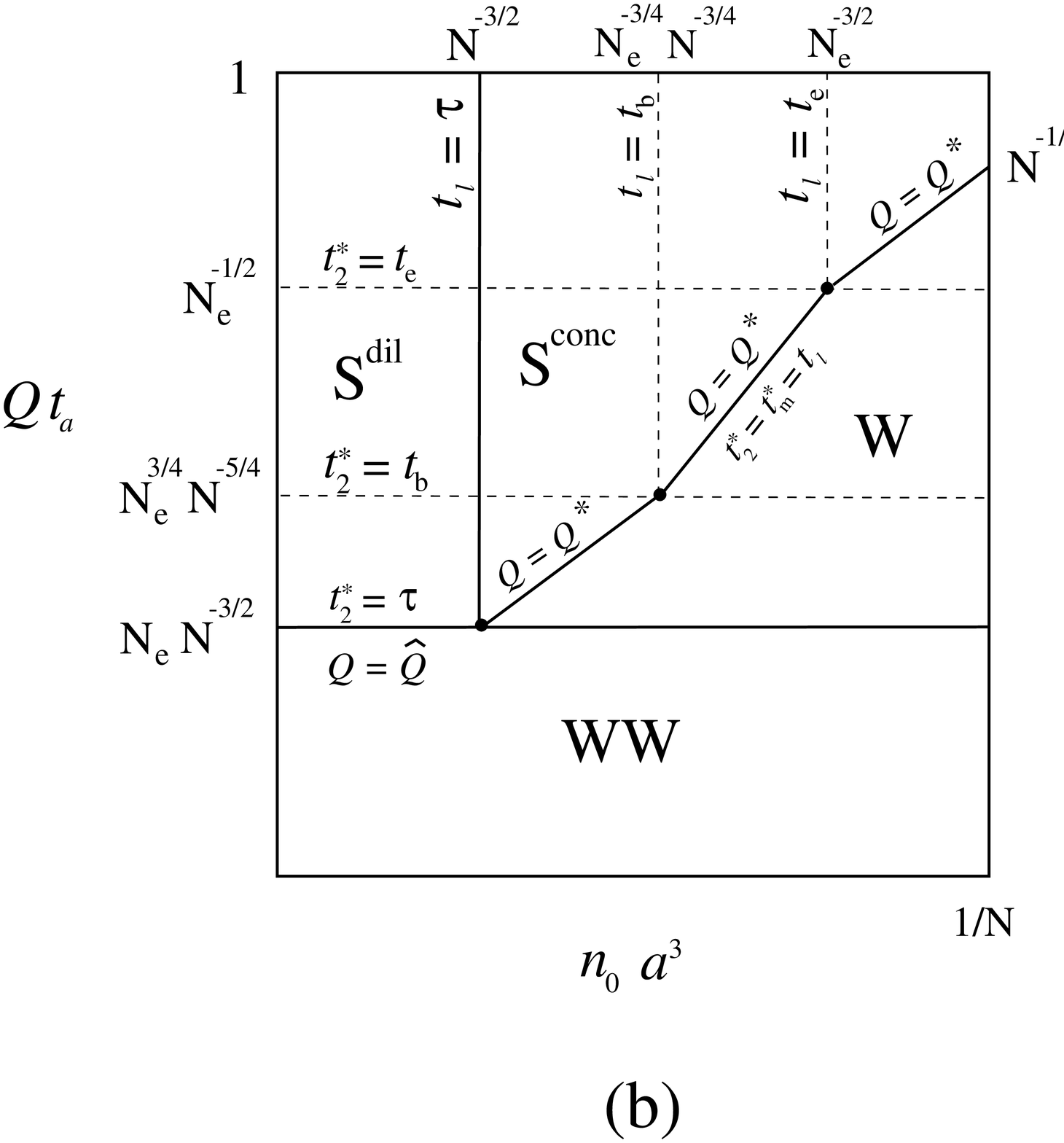}

\end{figure}

\mbox{\ }

\vfill

\mbox{\ } \hfill {\huge Fig.\@ \thefignumber (b)} 

\pagebreak


\end{document}